# Single-Switch Transformer-less Power Supply for Low Temperature Plasma Jet – 3.3 kV SiC MOSFET opportunities

David Florez, Hubert Piquet, Eric Bru, and Rafael Diez, *Senior Member, IEEE*

*Abstract*—**This work presents a simple power converter, without any high voltage transformer, able to supply and control a plasma jet based on dielectric barrier discharge. The converter, operating in pulsed current mode, requires a single power switch and is fed by a low voltage DC source. It can deliver very short duration pulses to the plasma jet with high current amplitude. The operating principle is explained by means of the state plane analysis and is validated with simulations and experimental results. The equations provided allow for the calculation during the design stage of important characteristics of the plasma jet as the peak voltage and the duration of the pulses. The power can be easily adjusted during experimentation to comply with the desired appearance of the plasma jet.**

## I. INTRODUCTION

A plasma jet could be seen as a source of low-temperature excited species sent to a target via a spray (or stream) without a necessary physical contact between the supply and the target. For biomedical applications, this target might be a human tissue to be treated in curing applications [1] [2] [3] [4] [5] where low temperature is mandatory to avoid pain in the patient. This source of non-equilibrium plasma is also matter of research for other usages where a temperature near the ambient must be maintained, such as in material processing [6] [7], enzymes production [8] [9], water depollution [10] [11], and agricultural production [12] [13].

A dielectric barrier discharge (DBD) plasma jet is obtained by injecting a gas flow, typically He or Ar [14] [15], commonly selected for their low breakdown electric field, into one side of an insulating pipe (plastic, glass or quartz). On this dielectric cylinder, two ring electrodes are disposed, as shown in Fig. 1. Those metallic electrodes are supplied by a high-voltage generator that creates a DBD in the volume between the rings, and the flow exits in the other side of the pipe (the nozzle) forming the jet of excited species (the plume).

The low temperature is maintained thanks to the DBD arrangement that avoids the increase of the current density and the transition to the arc regime in the gas [16] [17]. The discharge is produced inside the pipe because the gas flowing presents a lower breakdown voltage than the air outside the metallic rings, that are also covered by an insulator [18] (silicone paste in Fig. 1c).

The power supplies found in the literature for plasma jets are the same as for conventional DBDs, mainly of pulsed [19] [20] [21] [22] [23] [24] or sinusoidal [25] [26] [27] [28] [29] voltage waveform, including typically a step-up transformer and at least two switches [30] [31] [32] [33] [34] [35] [36]. According to previous investigations [5], the best performances are obtained when very sharp and high intensity current pulses (below 1 μs) are injected into the plasma; the required voltage levels are in the 3 kV range with a power below 5 W. In Table I the main characteristics of existing topologies are compared.

Table I. Plasma jet power supplies in the literature

| Main output features | [Ref], Topology if reported | Disadvantages | Peak current | Current pulse duration |
|---|---|---|---|---|
| Square voltage | [19] Topology not reported | Hard switching, high voltage (HV) transformer | 40 mA | ~700 ns |
| Square voltage  No HV transformer | [34], [36] Half-Bridge inverter + HV transformer + voltage multiplier + series of several low voltage switches | Many stages and components, hard switching and synchronization of low voltage MOSFETS | 2.6 A | ~ 80 ns |
| Sine voltage | [25] Topology not reported | HV transformer, not portable | ~ 60 mA | Mainly sinusoidal (not pulsed) |
| Sine voltage | [28] Commercial, Audio Frequency generator. | Low-frequency and HV transformer. not portable | ~ 150 mA | |
| Sine voltage | [29] Commercial, Topology not reported. | Not easy adjustment of the jet power | ~ 60 mA | |
| Sine current source | [35] Full-Bridge + HV Transformer | Many switches, high voltage transformer | ~ 35 mA | |
| Pulsed current | [5] Series resonant inverter | High voltage transformer, many switches | ~ 150 mA | ~ 120 ns |

The topologies with high-voltage transformer connecting the plasma jet are limited when trying to obtain short current pulses

This work was jointly supported by LAPLACE laboratory and Pontificia Universidad Javeriana, Project 10035.

*(Corresponding author: David Florez)*. David Florez and Rafael Diez are with the Department of Electronics Engineering, Pontificia Universidad Javeriana, Bogotá 110231, Colombia (e-mail: d.florez@javeriana.edu.co, rdiez@javeriana.edu.co).

Hubert Piquet and Eric Bru are with LAPLACE Laboratory, Université de Toulouse, CNRS, 31071 Toulouse, France (e-mail: hubert.piquet@laplace.univ-tlse.fr, eric.bru@laplace.univ-tlse.fr).

in the discharge because of the leakage inductance and output capacitance of this transformer. Authors in [34] [36] obtain a good performance in this aspect placing a series connection of several switches in the output. Nevertheless, this solution is complex and not as portable as desired for low power and medical applications where a compact device is an advantage.

The voltage mode supplies (pulsed and sinusoidal) tend to present output power highly dependent on external operating conditions (*e.g.* temperature) due to changes on $dv/dt$ waveform characteristics and the capacitive nature of the load, as will be explained in section II. This issue must be addressed on the converter design for medical applications where a stable and controllable output power is necessary.

In the present work, a power supply is proposed focused on a plasma jet for medical treatment: typical flow of He is about 1 L/min, with an inner pipe diameter of 3 mm. A transformer-less single switch converter configuration is conceived taking advantage of the recent 3.3 kV SiC semiconductor technology that produced commercial high voltage devices that now are compatible with the plasma jet voltage rating. Therefore, a soft-switching topology based on the energy transfer of the buck-boost operating principle is studied and experimentally verified.

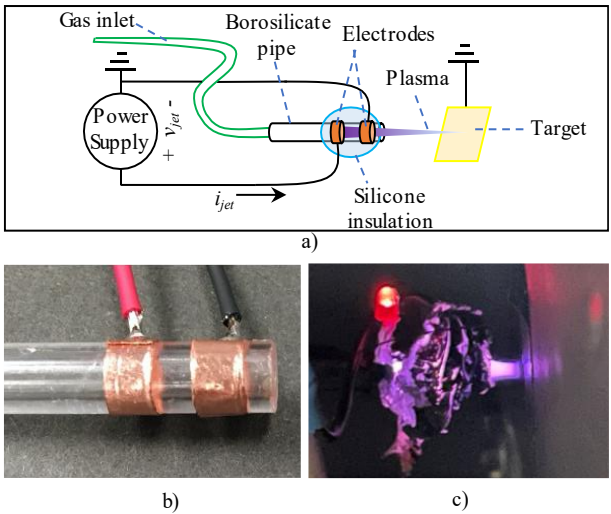

**Fig. 1.** DBD plasma jet. Components and connections (a). Configuration of the electrodes before insulation (b). Plasma jet ignited getting out of the pipe, an LED is included only for scale purposes – size: 3mm (c).

First, a basic explanation of the electrical model for a plasma jet to incorporate this circuit elements into the power supply design is presented. Then, the operating principle of the converter is explained thanks to the state plane analysis obtaining the dimensioning equations. The theoretical results are finally compared with simulations and experimental results.

## II. Modeling and Supply of Plasma Jets

In this section a simple equivalent electric circuit is obtained for the plasma jet reactor used in this work as shown in Fig. 2. Ideally, the path of the current, $i_{jet}$, that creates the low-temperature plasma could be seen starting from the positive terminal of the power supply and its connection with one of the ring electrodes, then crossing the dielectric barrier of the pipe wall to reach the gas. This is represented by $C_{d1}$ in Fig 2(b). The current in the gas could be capacitive or conductive depending on the state of the gas. If the gas voltage, $v_g$, is less than the breakdown voltage, the gas is an isolator represented by $C_g$. If the gas voltage increases and produces the breakdown inside the pipe, a low temperature plasma is created, and the gas voltage could be considered as constant, $\pm$ $V_{th}$, with the sign being the same as for $i_{jet}$. Finally, the current crosses again the dielectric barrier to attain the second ring electrode and the negative terminal of the power supply, that is usually connected to the ground. This capacitive connection is represented by $C_{d2}$.

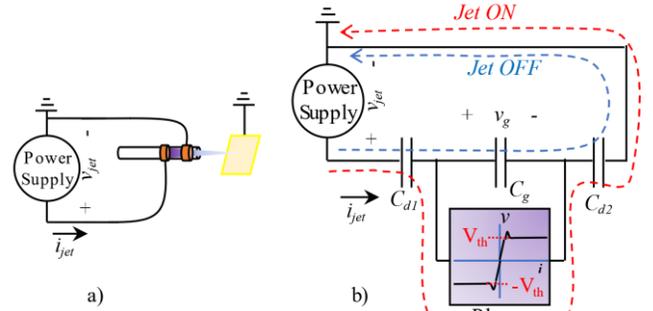

**Fig. 2.** Electric model of the plasma jet. Physical connections (a) and the simplified equivalent electric circuit (b).

The identification of the parameters for the equivalent circuit is done using the Manley diagram method, as detailed in [5], obtaining for the experimental arrangement used in this work: $C_g = 11$ pF, $C_d = 38$ pF, $V_{th} = 470$ V; being $C_d$ the series equivalent capacitance of $C_{d1}$ and $C_{d2}$. Note that the jet voltage, $v_{jet} = v_g + v_{Cd}$ , will reach higher values than the gas voltage, in the order of kV.

## III. Single Switch Transformer-less Topology

The converter topology proposed in this work can be seen in Fig. 3. It consists only of a DC power supply ($V_{in}$), an inductor ($L$) and, a switch implemented with a MOSFET ($S$); its body diode takes part in the operation. A 3.3 kV SiC MOSFET, able to sustain the jet voltage, is selected due to its short switching times, especially during turn-off: this is necessary in order to obtain current pulses with duration below microseconds. The low value for the output capacitance of this SiC MOSFET, compared to other semiconductor devices, is also advantageous to improve the energy transfer to the plasma jet.

In this figure, the $C_t$ capacitor is introduced to gather several parasitic elements: the stray capacitance of the inductance, the drain-source and drain-gate capacitances of the MOSFET, the external capacitance between one of the ring electrodes and the ground of the DBD jet, and the capacitances of the probes that are placed to take measurements. This equivalent capacitance is in the same order of magnitude as the gas and dielectric capacitances and must be considered within the operating principle of the converter. Note that the gas model is the same as in Fig. 2. However, the circuit notation with a diode bridge and a constant voltage $V_{th}$ is used to build the equivalent circuits further in this chapter.

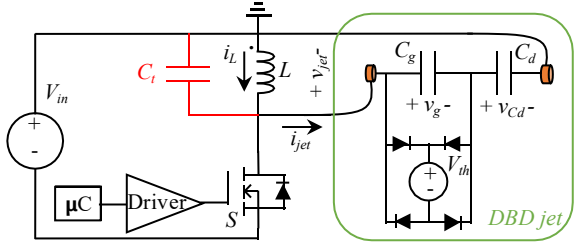

**Fig. 3.** Single switch transformer-less converter, considering a global parasitic capacitance to be included in the analysis.

### A. Operating Principle

The operating principle of the converter is based on energy storage in the inductance while $S$ is ON. Part of this energy is delivered later to the load when $S$ turns-off, and part is returned to the DC input power supply, considering that the inductance current $i_L$ is bidirectional.

Next, a five-steps sequence shown in Fig. 4, will be detailed with the help of the state plane analysis [37] [38] and assuming that, an initial positive current in the inductance is present ($i_L = I_{L0}$):

#### 1) **Positive Jet current before breakdown** ($t1 \leq t \leq t2$)

The equivalent circuit for the first state, starting at t1 when the S switch is turned OFF, is presented in Fig. 4a, configuring a resonant circuit between the inductance $L$, and $C_l$ that is the equivalent parallel capacitance of, $C_t$ and $C_{eq}$, being the latter the series equivalent capacitance of $C_d$ and $c_g$:

$$C_1 = C_t + C_{eq} = C_t + \frac{c_d \cdot c_g}{c_d + c_g} \qquad (1)$$

The state plane circular trajectory between $t1$ and $t2$, shown in Fig. 4f that is centered in the origin, describes the evolution of the inductance current *vs.* the voltage of the equivalent capacitance (the total jet voltage), with the normalization as follows:

$$R_{O1} = \sqrt{\frac{L}{C_1}} \qquad (2)$$

$$\omega_{O1} = \sqrt{\frac{1}{L \cdot C_1}} \qquad (3)$$

$$j_L = i_L \cdot \frac{R_{O1}}{V_{in}} \qquad (4)$$

$$m_{jet} = \frac{v_{jet}}{V_{in}} \qquad (5)$$

If the initial jet voltage, $v_{jet}$, is assumed equal to -$V_{in}$, or in normalized terms $m_{jet}$ = -1, the jet voltage will increase, and so the gas voltage, $v_g$, until the breakdown happens when the gas voltage, $v_g$, attains +$V_{th}$ at $t = t_2$, igniting the jet and changing the equivalent circuit to the one in Fig. 4b.

During the [$t_1$, $t_2$] subinterval, $v_g$ starts at -Vth and rises to +Vth, so the change of charge in this capacitor is equal to:

$$\Delta Q_g = C_g \cdot 2 \cdot V_{th} \qquad (6)$$

The change of voltage in the jet can be obtained using the same charge variation in $C_{eq}$. Taking also the initial jet voltage, the normalized jet voltage ($m_{jet}$) at the time of this first breakdown, $t_2$, can be obtained as:

$$M_{br1} = m_{jet}(t_2) = -1 + \frac{2 \cdot M_{th} \cdot (c_d + c_g)}{c_d} \qquad (7)$$

And using Pythagoras' theorem, the inductance current at the same instant ($t_2$) can be found as:

$$J_{br1} = j_L(t_2) = \sqrt{1 + J_{Lo}^2 - M_{br1}^2} \qquad (8)$$

These current and voltage values are used as the initial conditions for the next subinterval.

#### 2) **Positive Discharge Current** ($t2 \leq t \leq t3$)

Now, with the jet ignited and using the equivalent circuit shown in Fig. 4b, the state plane trajectory between $t_2$ and $t_3$ is traced in Fig. 4g. Once the breakdown is established, the power delivery takes place. Its value is the product of the $V_{th}$ gas voltage with the jet current $i_{jet}$. The time intervals when energy is injected into the plasma are [$t2,t3$] as well as [$t4,t5$] (subsection 4). As the equivalent capacitance during this sequence is different, a new normalization is used for the state plane, now tracing the inductance current *vs.* the voltage of the $C_2$ equivalent capacitance:

$$R_{O2} = \sqrt{\frac{L}{C_2}} \qquad (9)$$

$$\omega_{O2} = \sqrt{\frac{1}{L \cdot C_2}} \qquad (10)$$

Being

$$C_2 = C_t + C_d \qquad (11)$$

$$y_L = i_L \cdot \frac{R_{O2}}{V_{in}} = j_L \cdot \sqrt{\frac{C_1}{C_2}} \qquad (12)$$

$$m_{C2} = \frac{v_{C2}}{V_{in}} \qquad (13)$$

Note that the Thévenin's equivalent voltage of the resonant circuit is:

$$V_{thp} = V_{th} \frac{c_d}{c_d + c_t} \quad (14)$$

And the total jet voltage in this subinterval can be calculated as:

$$v_{jet} = V_{thp} + v_{C2} \quad (15)$$

This trajectory, centered at $m_{C2} = -M_{thp}$, is maintained until the plasma jet is extinguished when the inductance current crosses by zero. At this $t_3$ instant, the equivalent $C_{p2}$ voltage can be calculated as:

$$m_{c2}(t_3) = -M_{thp} + r2 \quad (16)$$

With:

$$r2 = \sqrt{Y_{br1}^2 + M_{br1}^2} \quad (17)$$

### 3) Second Ignition with Negative Current ($t3 \leq t \leq t4$)

The current in the inductance will continue flowing with a negative sign and the equivalent circuit of Fig. 4c, returning to the state plane of Fig. 4f, and the normalization used in the first subinterval. This new trajectory, also centered in the origin, starts with zero current and with the peak value of the jet voltage ($\widehat{M}_{jet}$).

Here, the gas voltage starts at $+V_{th}$ and goes down to $-V_{th}$, to create a new discharge but with a negative inductance current. Therefore, the change of charge in this capacitor is equal to:

$$\Delta Q_g = -C_g \cdot 2 \cdot V_{th} \quad (18)$$

The change of voltage in the jet can be obtained again using the same charge variation in $C_{eq}$. Considering that this sequence starts with the peak voltage in the jet, the normalized jet voltage at the instant of the second breakdown, $t_4$, can be obtained as:

$$M_{br2} = m_{jet}(t_4) = \widehat{M}_{jet} - \frac{2 \cdot M_{th} \cdot (C_d + C_g)}{C_d} \quad (19)$$

And the normalized inductance current, at the same instant, as:

$$J_{br2} = j_L(t_4) = -\sqrt{\widehat{M}_{jet}^2 - M_{br2}^2} \quad (20)$$

### 4) Negative Discharge Current ($t4 \leq t \leq t5$)

A new discharge is established at $t_4$, with a negative inductance current, ruled by the equivalent circuit in Fig. 4d. The state plane and normalization of Fig. 4g must be used. Noting that the center of the circular trajectory has changed to $+X_{thp}$, differing in sign from the positive discharge current sequence.

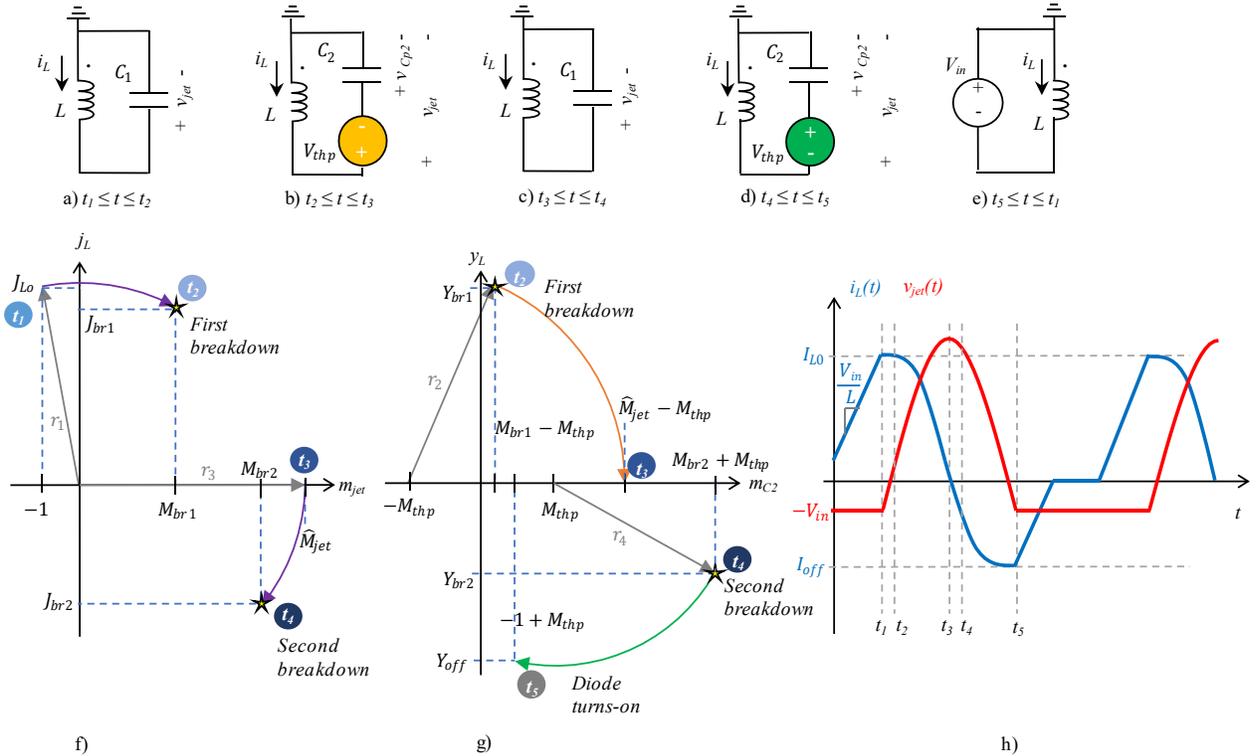

**Fig. 4.** Operating principle of the converter. Equivalent circuits for different stages: reaching the gas breakdown (a) and (c). energy injection to the plasma jet (b) and (d). Recovery of the initial inductance current (e). State plane for first and third subintervals (f) and for the second and fourth subintervals (g). Waveforms for the inductance current and the jet voltage (h).

In this sequence the total jet voltage can be calculated as:

$$v_{jet} = v_{C2} - V_{thp} \qquad (21)$$

The current delivery from the inductance to the jet finishes when the body diode of the MOSFET turns-on, making constant the voltage across all the capacitors and their current equal to zero. This happens when:

$$m_{C2} = -1 + M_{thp} \qquad (22)$$

At this instant, $t_5$, the normalized inductance current can be computed as:

$$Y_{off} = y_L(t_5) = -\sqrt{r_4^2 - 1} \qquad (23)$$

with

$$r4 = \sqrt{Y_{br2}^2 + M_{br2}^2} \qquad (24)$$

and

$$Y_{br2} = J_{br2} \cdot \sqrt{\frac{c_1}{c_2}} \qquad (25)$$

### 5) Inductance charging ($t5 \le t \le t1$)

After the body diode turns-on, the current in the inductance grows linearly because it is in parallel to $V_{in}$, as shown in Fig. 4e. There are two possibilities to regain the initial current in the inductance ($i_L = I_{L0}$) as shown in Fig. 5.

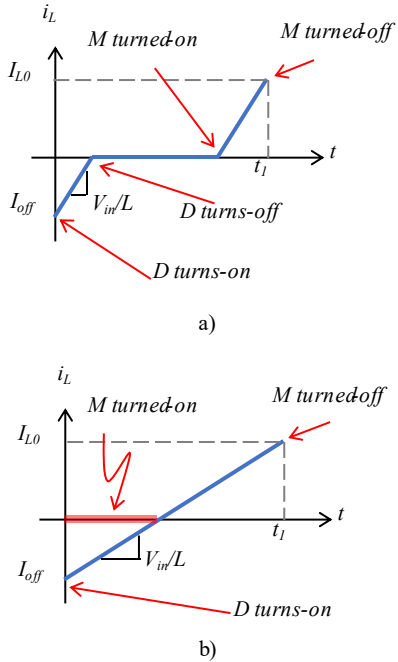

a)

b)

**Fig. 5.** Two options to regain the initial current in the inductance. Hard switching on the MOSFET turn-on (a). ZVS turn-on of the MOSFET (b).

The first one (Fig. 5a) has the possibility to maintain the inductance current equal to zero, once the diode spontaneously turns-off, before turning-on the MOSFET. This waiting time being a degree of freedom to adjust the plasma jet power via the switching frequency.

The other option (Fig. 5b) presents the advantage of zero voltage switching turn-on of the MOSFET before the diode turns-off. Nevertheless, in this configuration the switching frequency can be only slightly adjusted by the DC input voltage ($V_{in}$), almost losing the frequency as a parameter to adjust the plasma jet power.

Note that the MOSFET presents very low switching losses in both possibilities, with ZCS or ZVS at turn-on respectively. For the turn-off, at $t_1$, even if the current is high, the drain-source voltage increases slowly with the same rate as the jet voltage; therefore, it could be considered also as ZVS.

### B. Governing Equations

An important parameter for the plasma practitioner but also for the power supply designer, to avoid breakdown of the MOSFET, is the peak voltage of the jet, that can be written as:

$$\hat{V}_{jet} = \frac{\sqrt{I_{L0}^2 L + C_d V_{in}^2 + 4 C_d V_{th}^2 + 4 C_g V_{th}^2 + C_t V_{in}^2 - 4 C_d V_{in} V_{th}}}{\sqrt{C_d + C_t}} \qquad (26)$$

The average power delivered to the plasma jet is calculated with the difference of energy in the inductance between $t_1$ and $t_5$, multiplied by the switching frequency ($f$). Therefore, this power is a function of the initial and final currents of the inductance, as follows:

$$P_{jet} = \frac{1}{2} \cdot L \cdot \left(I_{L0}^2 - I_{off}^2\right) \cdot f \qquad (27)$$

The closed expression can be written as:

$$P_{jet} = f \left(2 C_d V_{th} \hat{V}_{jet} - 4 C_d V_{th}^2 + 4 C_g V_{th}^2 - 2 C_d V_{in} V_{th}\right) \qquad (28)$$

Note that this calculation does not consider the conduction losses of the passive components, neither losses on the MOSFET and its body diode.

Besides the power delivered to the plasma and the peak voltage of the jet, other important information is the duration of the pulses when the plasma is ignited, corresponding to the sequences of the two arcs of Fig. 4g. These time intervals can be calculated as:

Positive current pulse:

$$t_3 - t_2 = \frac{arctg\left(\frac{Y_{br1}}{M_{br1}}\right)}{\omega_{o2}} \qquad (29)$$

Negative current pulse:

$$t_5 - t_4 = \frac{\pi - arctg\left(\frac{Y_{br2}}{-M_{br2}}\right) - arctg(-Y_{off})}{\omega_{o2}} \qquad (30)$$

### C. Dimensioning

With the equations previously obtained, the DC input voltage, the inductance value, and the operating frequency are selected to comply with the desired power and the desired discharge pulse durations, maintaining the peak voltage in the safety range for the MOSFET.

The upper limit of the operating frequency is given by the inductance charge duration, $t_1$ in Fig 5b. The minimum frequency can be as low as zero thanks to the absence of transformer. In our case, the value of this frequency is experimentally adjusted to obtain the desired power in the plasma jet.

For example, with $V_{in} = 48$ V; $L = 28$ µH, $i_{L0} = 4.15$ A, and $f = 63$ kHz, the calculated peak voltage in the plasma jet is 2674 V. This value plus $V_{in}$ is well below the 3300 V that is the maximum rating of the MOSFET that will be used for the experimentation: G2R1000MT33J by GeneSic Semiconductor [39], commercially available only from the end of 2020. With these values the average power delivered by the DC power supply is 3.4 W, and duration of positive and negative discharge pulses of 50.7 ns and 32.7 ns respectively. These results are assuming a parasitic capacitance $C_t = 35$ pF.

Note that the proposed converter is very useful to obtain very short pulses with high current amplitude, which is desirable in several applications, and difficult to obtain with a typical converter with transformer due to the leakage inductance. It is important to maintain $C_t$ as low as possible to maximize the energy transfer as this capacitor is in parallel with the plasma jet. If $C_t$ increases, higher peak current in the inductance (and in the switch) and higher peak voltage will be necessary to obtain the same jet power.

The adjustment of the power during the experiments can be done by two methods. The first one is by changing the switching frequency (adjusting the waiting time mentioned in fig. 5a) which is completely linear. The second one by changing the current of the inductance at $t_1$ ($I_{L0}$) by means of the Vin voltage and/or by the time the inductance charges [$t_5$:$t_1$] subinterval.

## IV. VALIDATION OF THE OPERATING PRINCIPLE

### A. Simulations

At first, an ideal simulation in PSIM is performed to verify that the derived equations are correct. Results entirely agree with the values shown for the dimensioning example case. Subsequently, a different simulation is run on LTspice, using the MOSFET model provided by the manufacturer, and decreasing the external $C_t$ to 25 pF, because the MOSFET model already has its parasitic capacitances, with 10 pF for the output capacitance according to the manufacturer.

The inductance current and the jet voltage waveforms are plotted in Fig. 6 (top) the repetitive pulsed operation can be seen with very narrow voltage pulses for the jet voltage during the negative edge of the inductance current. A zoom to distinguish the different subintervals of the pulse, studied with the state plane analysis, is made in Fig. 6 (bottom). It is possible to identify the different times and the important values, as the peak voltage and duration of the positive and negative discharges. All in good accordance when compared with the theoretical expressions.

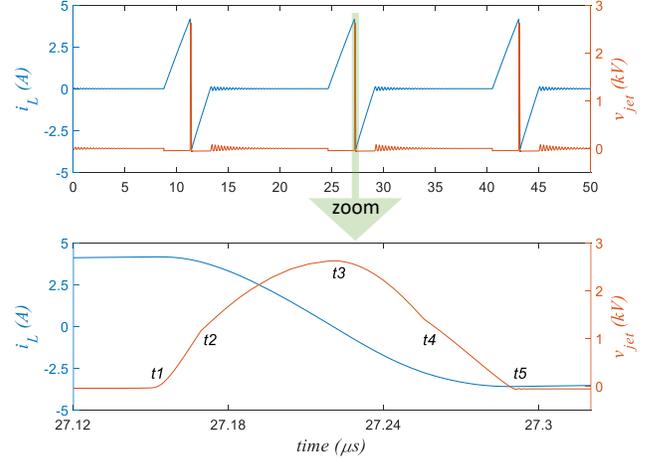

**Fig. 6.** Simulation of the inductance current and jet voltage waveforms. Three switching periods showing the charge of the inductance and the pulsed operation (top). Zoom during the energy transfer to identify the different subintervals studied by the state plane (bottom). $V_{in} = 48$ V; $L = 28$ µH, $i_{L0} = 4.15$ A, and $f = 63$ kHz.

### B. Experimental Results

Experiments are carried out with the same parameters used in the simulation. The implemented prototype and the obtained waveforms can be seen in Fig. 7 and Fig. 8, proving that the operating principle is respected when compared with simulation. These waveforms are acquired with a LeCroy HDO4024 oscilloscope, Testec TT-SI-9010A differential HV voltage probe, and LeCroy AP015 Current probe.

The experimental waveforms present some higher frequency oscillations during the pulse, because the electric model used to derive the equations is very simple and only considers two discharges per switching period. In fact, the plasma jet can present multiple discharges during this interval and a more precise electric model should be used to predict better the experimental waveforms.

Note that the experimental duration of the entire pulse (from $t_1$ to $t_5$) takes 115 ns against 132 ns predicted by the state plane theory, that could be considered a good approximation for this kind of application, where a very short output pulse could be obtained thanks to this transformer-less topology.

The experimental peak voltage of the jet is 2622 V, very close to the 2674 V predicted by the equations. A difference was found for the power delivered by the converter (5.8 W) much greater than the expected value (3.4 W), that can be attributed to the very low power and high intensity current of the application, making this calculation sensitive to the modeling of the plasma jet and the losses: skin effect, wiring and core losses, etc. This is to be studied in further work and could be explored with a more complex model of the plasma jet. Nevertheless, note that the value of the experimental power can be effectively adjusted by the power supply, by means of the switching frequency or $I_{L0}$ value, to comply with the desired appearance and characteristics of the plasma jet, as the distance or the temperature.

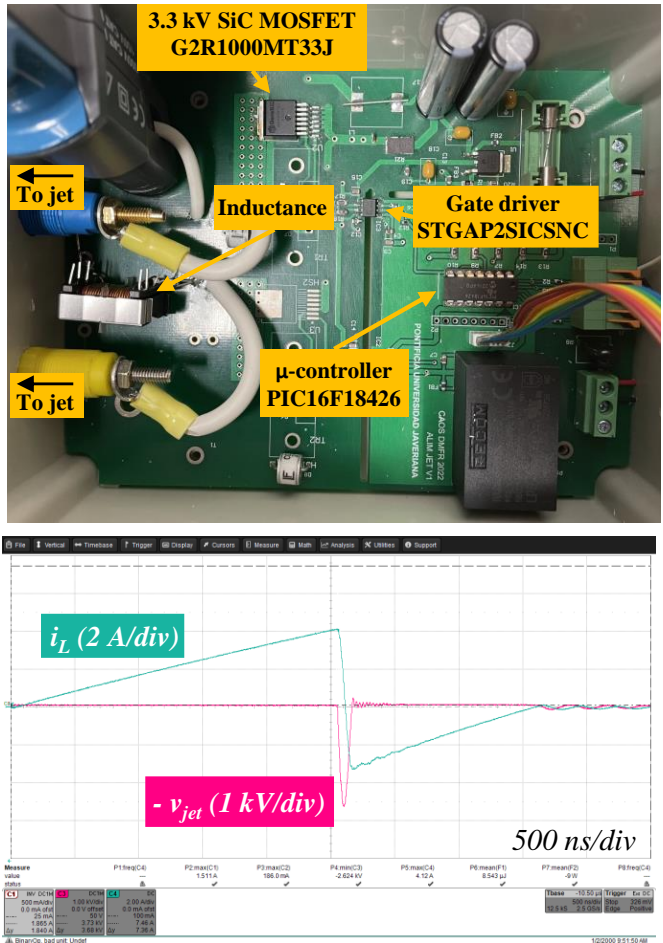

**Fig. 7.** Implementation of the power supply and experimental waveforms on oscilloscope. $V_{in} = 48$ V; $L = 28$ μH, $i_{L0} = 4.15$ A, and $f = 63$ kHz.

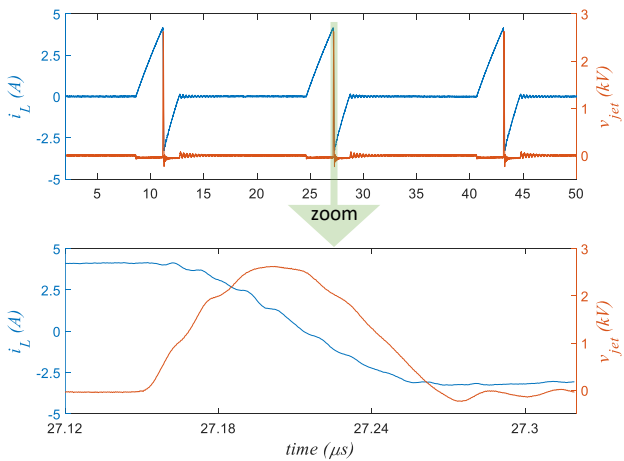

**Fig. 8.** Experimental waveforms complying with the operating principle. Periodic operation with the principle of inductance charging with waiting time (a). Zoom to verify the pulsed operation studied with the state plane (b). $V_{in} = 48$ V; $L = 28$ μH, $i_{L0} = 4.15$ A, and $f = 63$ kHz.

## V. CONCLUSION

The present work presents a very compact power supply for plasma jets using a single switch and without transformer. Besides the simplicity and low cost of the topology, it presents important output characteristics as the very short duration of the discharge pulses, as well as the high intensity current, that could lead to a very high efficiency of the plasma process.

This very simple topology has been possible thanks to the commercial 3.3 kV MOSFET technology developed in recent years. The low switching losses enables the use of the MOSFET without any heatsink for the usual lower power of the considered plasma jet application. Other important aspect of this power supply is the possibility to have a low value for the DC input voltage, as the 48 V used for experiments in this article.

The operating principle of this converter, studied with the state plane analysis, is validated with simulations and with experimental results. The equations derived are useful for the designer of the power supply to comply with the plasma practitioner requirements. The power delivered to the plasma jet can be easily adjusted, during experimentation, by means of the switching frequency or by changing the current in the inductance before the MOSFET turns-off.

This topology could be explored in the present with other types of DBD that present a low peak voltage, or in the future with a switch that increases even more its drain-source breakdown voltage. Further studies are also expected with a different 3300 V MOSFET, having lower drain-source resistance (50 mΩ instead of 1 Ω) and higher pulsed current rating (235 A instead of 10A), to increase the efficiency and reduce the duration of the pulses.

## REFERENCES

[1] B. Eggers, M. Stope, J. Marciniak, A. Mustea, J. Deschner, M. Nokhbehsaim and F.-J. Kramer, "Modulation of Inflammatory Responses by a Non-Invasive Physical Plasma Jet during Gingival Wound Healing," *Cells*, vol. 11, no. 2740, 2022, doi: 10.3390/cells11172740.

[2] J. Zhang, S. Peng, X. Zhang, R. Fan, X. Zhao, M. Qi, R. Liu, D. Xu and D. Liu, "Investigation of different solutions activated by air plasma jet and their anticancer effect," *Appl. Phys. Lett.*, vol. 120, no. 26, 2022, doi: 10.1063/5.0096605.

[3] A. Jo, H. Joh, J. Bae, S. Kim, T. Chung and et.al, "Plasma activated medium prepared by a bipolar microsecond-pulsed atmospheric pressure plasma jet array induces mitochondria-mediated apoptosis in human cervical cancer cells," *PLoS one*, vol. 8, no. e0272805, 2022, doi: 10.1371/journal.pone.0272805.

[4] X. Shi, S. Liu, R. Jiang, J. Chen, S. Jin, D. Mei, R. Zhou, Z. Fang and P. J. Cullen, "Development and characterization of touchable air plasma jet device for inactivation of oral bacteria,," *Results Phys.*, vol. 36, no. 105405, 2022, doi: 10.1016/j.rinp.2022.105405.

[5] A. Ivankov, T. Capela, V. Rueda, E. Bru, H. Piquet, D. Schitz, D. Florez and R. Diez, "Experimental Study of a Nonthermal DBD-Driven Plasma Jet System Using Different Supply Methods," *Plasma*, vol. 5, pp. 75-97, 2022, doi: 10.3390/plasma5010007.

[6] M. Shariat, "Plasma jet printing of silver nanoparticles on polyester fabric as a surface-enhanced Raman scattering substrate," *Optik*, vol. 277, no. 170698, 2023, doi: 10.1016/j.ijleo.2023.170698.

[7] R. Wang, Z. Xia, X. Kong, S. Xue and H. Wang, "Uniform deposition of silicon oxide film on cylindrical substrate by radially arranged

plasma jet array," *Surf. Coat. Technol.,* vol. 437, no. 128365, 2022, doi: 10.1016/j.surfcoat.2022.128365.

[8] H. L. Wapshott-Stehli, B. G. Myers, M. J. Herrera-Quesada, A. Grunden and K. Stapelmann, "Plasma-driven biocatalysis: In situ hydrogen peroxide production with an atmospheric pressure plasma jet increases the performance of OleTJE when compared to adding the same molar amount of hydrogen peroxide in bolus," *Plasma Process. Polym.,* vol. 19:e2100160, 2022, doi: 10.1002/ppap.202100160.

[9] A. Yayci, T. Dirks, F. Kogelheide, M. Alcalde, F. Hollmann, P. Awakowicz and J. E. Bandow, "Microscale Atmospheric Pressure Plasma Jet as a Source for Plasma-Driven Biocatalysis," *ChemCatChem,* vol. 12, no. 5893, 2020, doi: 10.1002/cctc.202001225.

[10] E. Abdel-Fattah, "Atmospheric pressure helium plasma jet and its applications to methylene blue degradation, *J Electrostat,* vol. 101, no. 103360, 2019, doi: 10.1016/j.elstat.2019.103360.

[11] H. E. Bousba, S. Sahli, W. Seif, E. Namous, L. Benterrouche and M. Saoudi, "Inactivation of Escherichia coli in water using cold atmospheric plasma jet," in *2nd International Conference on Advanced Electrical Engineering (ICAEE),* Constantine, Algeria, 2022, doi: 10.1109/ICAEE53772.2022.9962110.

[12] R. Kawakami, M. Aihara, T. Izumi, A. Shirai and T. Mukai, "Bactericidal effects of low-temperature atmospheric-pressure air plasma jets with no damage to plant nutrient solutions," *Biochem. Eng. J.,* vol. 187, no. 108661, 2022, doi: 10.1016/j.bej.2022.108661.

[13] Y. A. Ussenov, A. Akildinova, B. A. Kuanbaevich, K. A. Serikovna, M. Gabdullin, M. Dosbolayev, T. Daniyarov and T. Ramazanov, "The Effect of Non-Thermal Atmospheric Pressure Plasma Treatment of Wheat Seeds on Germination Parameters and α-Amylase Enzyme Activity," *IEEE Trans. Plasma Sci.,* vol. 50, no. 2, pp. 330-340, 2022, doi: 10.1109/TPS.2022.3145831.

[14] W. V. Gaens and A. Bogaerts, "Kinetic modelling for an atmospheric pressure argon plasma jet in humid air," *J. Phys. D: Appl. Phys.,* vol. 46, no. 275201, 2013, doi: 10.1088/0022-3727/46/27/275201.

[15] A. V. Nastuta and G. Torsten, "Cold Atmospheric Pressure Plasma Jet Operated in Ar and He: From Basic Plasma Properties to Vacuum Ultraviolet, Electric Field and Safety Thresholds Measurements in Plasma Medicine," *Appl. Sci.,* vol. 12(2), no. 644, 2022, doi: 10.3390/app12020644.

[16] F. Massines, G. Gouda, N. Gherardi, M. Duran and E. Croquesel, "The Role of Dielectric Barrier Discharge Atmosphere and Physics on Polypropylene Surface Treatment," *Plasmas and Polymers,* vol. 6, p. 35–49, 2001, doi: 10.1023/A:1011365306501.

[17] G. -M. Xu, Y. Ma and G. -J. Zhang, "DBD Plasma Jet in Atmospheric Pressure Argon," *IEEE Trans. Plasma Sci.,* vol. 36, no. 4, pp. 1352-1353, 2008, doi: 10.1109/TPS.2008.917772.

[18] G. -D. Wei, C. -S. Ren, M. -Y. Qian and Q. -Y. Nie, "Optical and Electrical Diagnostics of Cold Ar Atmospheric Pressure Plasma Jet Generated With a Simple DBD Configuration," *IEEE Trans. Plasma Sci.,* vol. 39, no. 9, pp. 1842-1848, 2011, doi: 10.1109/TPS.2011.2159810.

[19] S. Mashayekh, N. Cvetanović, G. B. Sretenović, Z. Liu, K. Yan and M. M. Kuraica, "Experimental study of a microsecond-pulsed cold plasma jet," *Eur. Phys. J. D,* vol. 77, no. 115, 2023, doi: 10.1140/epjd/s10053-023-00692-8.

[20] S. Jin, J. Chen, Z. Li, C. Zhang, Y. Zhao and Z. Fang, "Novel RDD Pulse Shaping Method for High-Power High-Voltage Pulse Current Power Supply in DBD Application," *IEEE Trans. Ind. Electron.,* vol. 69, no. 12, pp. 12653-12664, 2022, doi: 10.1109/TIE.2022.3140515.

[21] H. Gui, Z. Zhao, Q. Shi, X. Liu and C. Yao, "All-Solid-State Nanosecond Pulse Power Supply Based on BLTs and Pulse Transformer for DBD Application," *IEEE Trans. Power Electron.,* vol. 38, no. 8, pp. 10085-10092, 2023, doi: 10.1109/TPEL.2023.3274451.

[22] S. Jin, C. Zhang, Y. Peng and Z. Fang, "Novel IPOx Architecture for High-Voltage Microsecond Pulse Power Supply Using Energy Efficiency and Stability Model Design Method," *IEEE Trans. Power Electron.,* vol. 36, no. 9, pp. 10852-10865, 2021, doi: 10.1109/TPEL.2021.3064957.

[23] C. Sanabria, D. Florez, H. Piquet and R. Diez, "Sizing Equations for a Square Voltage Pulse Power Supply for Dielectric Barrier Discharges," *IEEE Trans. Power Electron.,* vol. 37, no. 4, pp. 4374-4384, 2022, doi: 10.1109/TPEL.2021.3122934.

[24] S. Jin and et.al, "A High-Drive-Performance Microsecond Pulse Power Module for Portable DBD Plasma Source Device," *IEEE Trans. Power Electron.,* vol. 38, no. 12, pp. 15072-15085, 2023, doi: 10.1109/TPEL.2023.3315524.

[25] O. Bastin, M. Thulliez, T. Serra, L. Nyssen, T. Fontaine, J. Devière, A. Delchambre, F. Reniers and A. Nonclercq, "Electrical equivalent model of a long dielectric barrier discharge plasma jet for endoscopy," *J. Phys. D: Appl. Phys.,* vol. 56, no. 125201, 2023, doi: 10.1088/1361-6463/acb603.

[26] X. Zhang, Y. Zhao and C. Yang, "Recent developments in thermal characteristics of surface dielectric barrier discharge plasma actuators driven by sinusoidal high-voltage power," *Chinese J. Aeronaut.,* vol. 36, no. 1, 2023, doi: 10.1016/j.cja.2022.01.026.

[27] M. Amjad, Z. Salam, M. Facta and S. Mekhilef, "Analysis and Implementation of Transformerless LCL Resonant Power Supply for Ozone Generation," *IEEE Trans. Power Electron.,* vol. 28, no. 2, pp. 650-660, 2013, doi: 10.1109/TPEL.2012.2202130.

[28] Y. D. Korolev, V. O. Nekhoroshev, O. B. Frants, A. V. Bolotov and N. V. Landl, "Power Supply for Generation of Low-Temperature Plasma Jets," *Russ. Phys. J.,* vol. 62, no. 11, pp. 2052–2058, 2020, doi: 10.1007/s11182-020-01944-5.

[29] S.-S. Huang, H.-C. Tsai, J. Chang, P.-C. Huang, Y.-C. Cheng and J.-S. Wu, "Establishing an equivalent circuit for a quasihomogeneous discharge atmospheric-pressure plasma jet with a breakdown-voltage-controlled breaker and power supply circuit," *J. Phys. D: Appl. Phys.,* vol. 55, no. 21, 2022, doi: 10.1088/1361-6463/ac4b57.

[30] M. Ponce-Silva, J. A. Aqui, V. H. Olivares-Peregrino and M. A. Oliver-Salazar, "Assessment of the Current-Source, Full-Bridge Inverter as Power Supply for Ozone Generators With High Power Factor in a Single Stage," *IEEE Trans. Power Electron.,* vol. 31, no. 12, pp. 8195-8204, 2016, doi: 10.1109/TPEL.2016.2520925.

[31] L. Chang, T. Guo, J. Liu, C. Zhang, Y. Deng and X. He, "Analysis and Design of a Current-Source CLCC Resonant Converter for DBD Applications," *IEEE Trans. Power Electron.,* vol. 29, no. 4, pp. 1610-1621, 2014, doi: 10.1109/TPEL.2013.2266376.

[32] Z. Deng, X. Liu, Q. Qiu, H. Jia, Y. Deng and X. He, "Nonlinear Resonance Characteristics Analysis of DBD Load Based on State Plane Trajectory," *IEEE Trans. Power Electron.,* vol. 38, no. 4, pp. 4760-4770, 2023, doi: 10.1109/TPEL.2022.3228786.

[33] S. Hao, X. Liu, W. Li, Y. Deng and X. He, "Energy Compression of Dielectric Barrier Discharge With Third Harmonic Circulating Current in Current-Fed Parallel-Series Resonant Converter," *IEEE Trans. Power Electron.,* vol. 31, no. 12, pp. 8528-8540, 2016, doi: 10.1109/TPEL.2016.2520953.

[34] S. Moshkunov, V. Khomich and E. Shershunova, "A High-Voltage Switching Power Supply for Cold Plasma Jets," *Tech. Phys. Lett.,* vol. 45, no. 2, pp. 93-95, 2019, doi: 10.1134/S1063785019020123.

[35] C.-F. Su, C.-T. Liu, J.-S. Wu and M.-T. Ho, "Development of a High-Power-Factor Power Supply for an Atmospheric-Pressure Plasma Jet," *Electronics,* vol. 10, no. 17, 2021, doi: 10.3390/electronics10172119.

[36] S. Moshkunov, N. Podguyko and E. Shershunova, "Compact high voltage pulse generator for DBD plasma jets," *J. Phys.: Conf. Ser.,* vol. 1115, no. 2, p. 022032, 2018, doi:10.1088/1742-6596/1115/2/022032.

[37] Y. Chéron, "Soft Commutation", London, U.K.: Chapman & Hall, 1992.

[38] R. Oruganti and F. C. Lee, "Resonant Power Processors, Part I---State Plane Analysis," *IEEE Trans. Ind. Appl.,* vol. IA21, no. 6, pp. 1453-1460, 1985, doi: 10.1109/TIA.1985.349602.

[39] GeneSic, "G2R1000MT33J datasheet," [Online]. Available: https://genesicsemi.com/sic-mosfet/G2R1000MT33J/G2R1000MT33J.pdf. [Accessed 30 October 2023].